**Disorder of Excitons and Trions in Monolayer MoSe$_2$**

Jue Wang[1], Christina Manolatou[2], Yusong Bai[1], James Hone[3], Farhan Rana[2,*], and X.-Y. Zhu[1,*]

[1] Department of Chemistry, Columbia University, New York, NY 10027, USA

[2] School of Electrical and Computer Engineering, Cornell University, Ithaca, NY 14853, USA

[3] Department of Mechanical Engineering, Columbia University, New York, NY 10027, USA

ABSTRACT. **The optical spectra of transition metal dichalcogenide (TMDC) monolayers are dominated by excitons and trions. Here we establish the dependences of these optical transitions on disorder from hyperspectral imaging of h-BN encapsulated monolayer MoSe$_2$. While both exciton and trion energies vary spatially, these two quantities are almost perfectly correlated, with spatial variation in the trion binding energy of only ~0.18 meV. In contrast, variation in the energy splitting between the two lowest energy exciton states is one order of magnitude larger at ~1.7 meV. Statistical analysis and theoretical modeling reveal that disorder results from dielectric and bandgap fluctuations, not electrostatic fluctuations. Our results shed light on disorder in high quality TMDC monolayers, its impact on optical transitions, and the many-body nature of excitons and trions.**

Transition metal dichalcogenide (TMDC) monolayers have emerged as the most versatile models for the exploration of many-body semiconductor physics in two-dimensions (2D). The interplay of 2D character and poorly screened Coulomb potential leads to strong many-body effects that dominate the optical properties of TMDCs. Strongly bound excitons, with binding energies in the hundreds of meV, and strongly bound trions, with binding energies in the tens of meV, have been observed in TMDCs [1–4]. These tightly bound excitonic complexes are attractive models for the understanding of many-body interactions in 2D and for optoelectronic applications. However, most 2D materials are strongly affected by disorder [5,6] whose sources include material defects [7], electrostatic potential fluctuations [8,9], dielectric constant variations [10], and strain-induced bandgap changes [11,12]. Interestingly, as we discuss in this paper, even state-of-the-art TMDC monolayers with hexagonal boron nitride (h-BN) encapsulation are not immune to disorder. The response of many-body exciton and trion states to disorder can provide valuable insights into both the nature of disorder and the nature of the excitonic complexes. Here we carry out hyperspectral imaging of excitons and trions in h-BN encapsulated monolayer MoSe$_2$. We find that both exciton and trion energies are sensitive to variations in the local environment, but these two energies are almost perfectly correlated, in contrast to the behavior of energy gap between the two lowest excitons states. Statistical analysis of the spatial energy variations, combined with theoretical modeling of exciton and trion states in the presence of disorder, reveals that the sources of disorder are dielectric constant and electronic bandgap variations of Δε~0.08 and ΔE$_g$~2-3 meV, respectively.

* Correspondence. email: xyzhu@columbia.edu (xyz); farhan.rana@cornell.edu (FR).

In experiments, we use the highest quality monolayers exfoliated from flux grown MoSe$_2$ single crystals with low defect density (~ $10^{11}$ cm$^{-2}$) [7] and large areas (>160 μm$^2$). Each MoSe$_2$ monolayer is encapsulated in h-BN. The steady-state photoluminescence (PL) spectra (Fig. 1a) show two narrow peaks assigned to the lowest energy *1s* exciton and the lowest energy trion [3,4], with mean energies $E_{1s}^{ex}$ = 1.6465 ± 0.0001 eV and $E^{tr}$ = 1.6214 ± 0.0001 eV, respectively. The difference $E_{1s}^{ex} - E^{tr}$ is often referred to as the trion binding energy $E_b^{tr}$. The high quality of the monolayer gives rise to near unity total PL quantum yield [13] with full-width-at-half-maximum (FWHM) of both exciton and trion approaching ~1 meV [7,14]. We determine the effects of disorder [7–12,15–18] by hyperspectral PL imaging with a spatial resolution of ~1 μm. A continuous wave laser at hν = 2.33 eV excites the sample in a diffraction-limited spot of FWHM~0.43 μm, at 4.66 μW/μm$^2$. We extract $E_{1s}^{ex}$ and $E^{tr}$ for each spot from intensity-weighted averaging of the PL spectra (Fig. 1a). While the PL intensity is relatively homogeneous (Fig. 1a, inset), $E_{1s}^{ex}$ and $E^{tr}$ fluctuate over the whole sample area (Fig. 1b,c). The difference between the two shows a surprisingly uniform spatial distribution with a mean value of $E_b^{tr}$ =26.220(1) ±0.0005 meV (Fig. 1d). This suggests that spatial fluctuations in $E_{1s}^{ex}$ and $E^{tr}$ are highly correlated. Fig. 1e shows a scatter plot of $E^{tr}$ vs. $E_{1s}^{ex}$. The solid and dashed lines indicate the directions of the eigenvectors of the covariance matrix of $E_{1s}^{ex}$ and $E^{tr}$ with slopes of +0.97 and -1.03, respective. Note that slopes of ±1 would indicate perfect correlation and slopes of 0 and ∞ would indicate no correlation. we plot histograms of $E_{1s}^{ex}$, $E^{tr}$, and $E_b^{tr}$ in Fig. 1f-h, corresponding to standard deviations of $\sigma_{1s}^{ex} = 3.6 \pm 0.2$ meV, $\sigma^{tr} = 3.5 \pm 0.2$ meV, and $\sigma_b^{tr} = 0.186 \pm 0.001$ meV, respectively. Because of the nearly perfect correlation between $E_{1s}^{ex}$ and $E^{tr}$, $\sigma_b^{tr}$ is only ~5% of $\sigma_{1s}^{ex}$ and $\sigma^{tr}$. The insensitivity of $E_b^{tr}$ to disorder is seen in a broad temperature range until T ~ 60 K, above which the trion PL peak disappears, likely attributed to dissociation of the many-body trion complex by phonon scattering (Fig. S1 and Fig. S2).

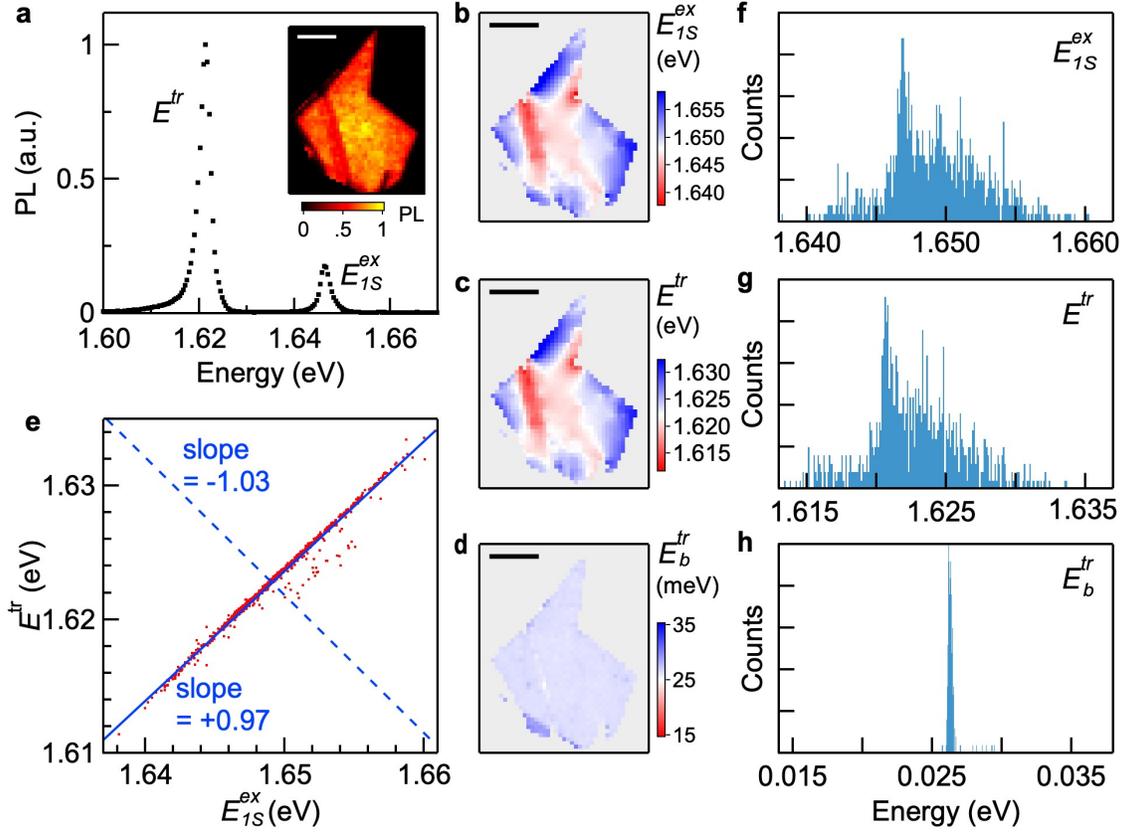

**Fig.1 Insensitivity of the trion binding energy to disorder in monolayer MoSe$_2$. a**, Representative PL spectrum of monolayer MoSe$_2$ encapsulated in h-BN. Inset: total PL intensity image. Scale bar: 5 μm. **b-d**, Spatially resolved exciton peak energy $E_{1s}^{ex}$ (**b**), trion peak energy $E^{tr}$ (**b**), and trion binding energy $E_b^{tr} = E_{1s}^{ex} - E^{tr}$ (**d**) extracted from the PL map. The ranges of color scales are 20 meV. Scale bar: 5 μm. **e**, Correlation between trion and exciton peak energies with data points (red dots) extracted from PL mapping. The solid and dashed blue lines indicate the directions of the eigenvectors of the covariance matrix and have slopes of +0.97 and -1.03, respectively. **f-h**, Histograms of $E_{1s}^{ex}$ (**f**), $E^{tr}$ (**g**) and $E_b^{tr}$ (**h**). The standard deviations in the data is $\sigma_{1s}^{ex} = 3.6 \pm 0.2$ meV (**f**), $\sigma^{tr} = 3.5 \pm 0.2$ meV (**g**) and $\sigma_b^{tr} = 0.186 \pm 0.001$ meV (**h**), respectively. All data shown is taken at a temperature of 4K.

In stark contrast to the nearly constant $E_b^{tr}$, the energy splitting between the exciton levels is more broadly distributed. We quantify the spatial distributions in *1s* and *2s* exciton energies from reflectance contrast ($R_c$) spectra. Fig. 2a is a representative $R_c$ spectrum showing the A-exciton *1s* (1.649 eV) and *2s* (1.799 eV, inset) transitions, and the B-exciton *1s* transition (1.847 eV), with energies in agreement with previous reports [10,19]. The $E_{1s}^{ex}$ and $E_{2s}^{ex}$ energy maps for the A-*1s* exciton in Figs. 2b and 2c show characteristic spatial variations attributed to disorder. The energy gap, $E_\Delta^{ex} = E_{2s}^{ex} - E_{1s}^{ex}$, also shows spatial variation of the same order (Fig. 2d). A scatter plot of $E_{2s}^{ex}$ vs $E_{1s}^{ex}$ is shown in Fig. 2e, along with solid and dashed blue lines showing the directions of the eigenvectors of their covariance matrix. The slopes, +1.44 and -0.69, indicate a much weaker correlation than that between $E_{1s}^{ex}$ and $E^{tr}$. Fig. 2f-h show histograms of $E_{1s}^{ex}$ and $E_{2s}^{ex}$, and $E_\Delta^{ex}$ with standard deviations $\sigma_{1s}^{ex} = 3.1 \pm 0.2$ meV, $\sigma_{2s}^{ex} = 4.4 \pm 0.4$ meV, and $\sigma_{\Delta}^{ex} = 1.7 \pm 0.1$ meV, respectively. The weak correlation between $E_{1s}^{ex}$ and $E_{2s}^{ex}$ results in $\sigma_{\Delta}^{ex}$ being 40-60% of $\sigma_{2s}^{ex}$ and $\sigma_{1s}^{ex}$.

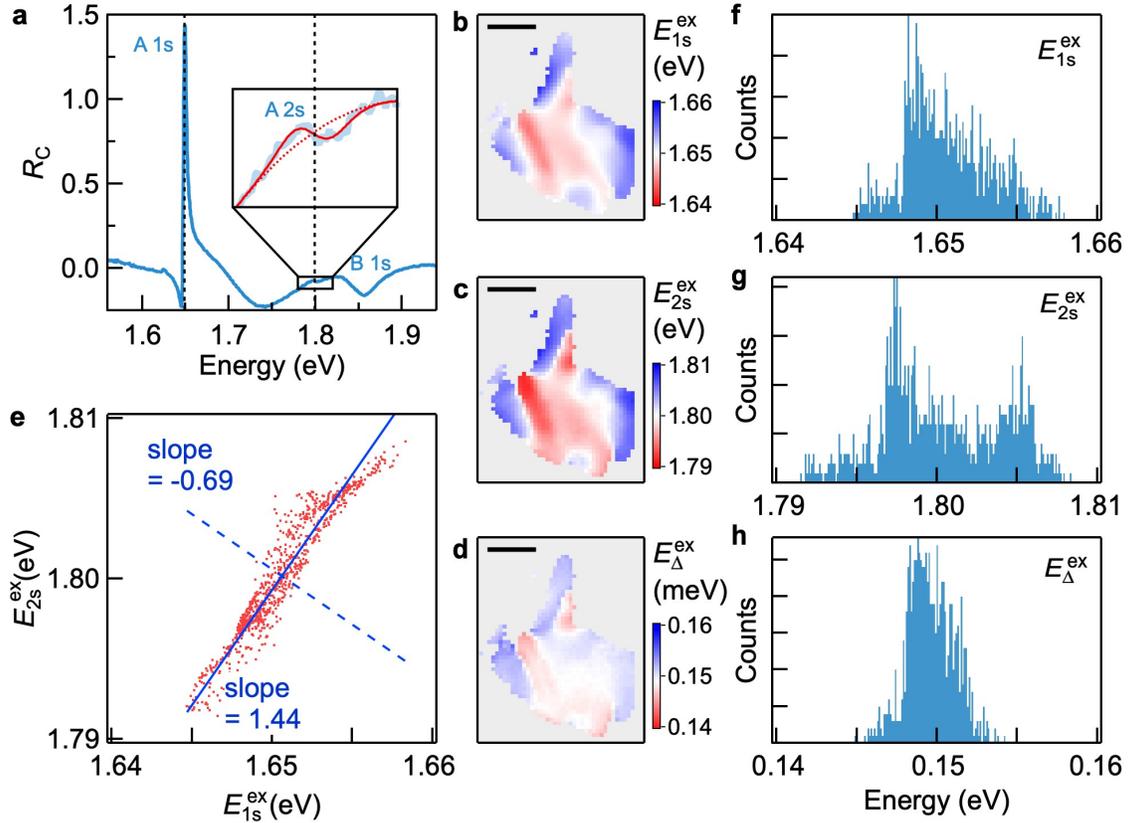

**Fig. 2. Spatial fluctuation of the exciton energy level splitting in monolayer MoSe$_2$. a.** A *1s* (1.649 eV), A *2s* (1.799 eV) and B *1s* resonances. Inset: Fit (red line) to $R_C$ around the A *2s* resonance (light blue line). **b-d,** Spatially resolved exciton peak energies $E_{1s}^{ex}$ (**b**), $E_{2s}^{ex}$ (**c**) and $E_\Delta^{ex} = E_{2s}^{ex} - E_{1s}^{ex}$, (**d**) extracted from $R_C$ spectral image. All three color scales span 20 meV. Scale bar: 5 μm. **e,** Correlation between $E_{2s}^{ex}$ and $E_{1s}^{ex}$ exciton energies (red dots). The solid and dashed blue lines indicate the directions of the eigenvectors of the covariance matrix with slopes of +1.44 and -0.69, respectively. **f-h,** Histograms of the exciton energies $E_{1s}^{ex}$ (**f**), $E_{2s}^{ex}$ (**g**) and the energy difference $E_\Delta^{ex}$ (**h**). The respective standard deviations are 3.08 ± 0.05 meV, 4.37 ± 0.05 meV and 1.73 ± 0.05 meV. The mean value of *2s-1s* energy difference $E_\Delta^{ex}$ is 149.46 ± 0.05 meV. All data shown is taken at a temperature of 4K.

The covariance matrices $K^a(E_{1s}^{ex}, E_{2s}^{ex})$ and $K^b(E_{1s}^{ex}, E^{tr})$ can be used to understand the nature of the disorder. These matrices are obtained from the data (in units of meV$^2$),

$$K^a(E_{1s}^{ex}, E_{2s}^{ex}) = \begin{bmatrix} 9.4763 & 12.7781 \\ 12.7781 & 19.0697 \end{bmatrix}, K^b(E_{1s}^{ex}, E^{tr}) = \begin{bmatrix} 12.8235 & 12.4329 \\ 12.4329 & 12.0843 \end{bmatrix} \quad (1)$$

The determinants of both covariance matrices are non-zero, *implying that more than one disorder mechanism is responsible for the observed spatial variations in the exciton and trion energies.* We consider the effects of two different types of spatial disorder on $E_{\Delta}^{ex}$ and $E_b^{tr}$: i) electronic bandgap variations due to strain [11,12], and ii) disorder in the dielectric constant of the media surrounding the 2D monolayer [10]. In the Supplementary Material, we discuss potential disorder and explain why it is inconsistent with our experimental observations.

Recent many body models have shown that the PL peaks observed in the measured optical spectra correspond to a superposition of exciton and trion states [20], also called exciton-polaron states [21–23]), rather than to pure exciton or pure trion states. Furthermore, the trions states involved in this superposition are 4-body neutral states [20] and not 3-body charged states, as is commonly assumed. However, given the small electron density in our samples ($\leq 10^{11}$ cm$^{-2}$), one can safely assume, in light of the model of Rana *et al.* [20], that the observed lowest energy exciton-trion superposition state in our PL spectrum is essentially a 4-body bound trion state ($E_{1s1s}^{tr}$) and the higher energy superposition states in PL and R$_c$ spectra are essentially 2-body bound exciton states ($E_{1s}^{ex}$ and $E_{2s}^{ex}$). If $\vec{R}$ is the center of mass coordinate of the exciton (or trion), the local shifts in the exciton and trion energies can be written as,

$$\Delta E_{ns}^{ex}(\vec{R}) = \Delta E_g^{S,T}(\vec{R}) + \left(\gamma_g + \gamma_{b-ns}^{ex}\right)\Delta\varepsilon_{ext}(\vec{R})$$
$$\Delta E^{tr}(\vec{R}) = \Delta E_g^{S,T}(\vec{R}) + \left(\gamma_g + \gamma_{b-1s}^{ex} + \gamma_b^{tr}\right)\Delta\varepsilon_{ext}(\vec{R}) \quad (2)$$

Here, $\Delta E_g^{S,T}$ is the variation in the bandgap attributed to strain, the coefficient $\gamma_g = \partial E_g / \partial \varepsilon_{ext}$ describes the change in the bandgap due to dielectric disorder, the coefficients $\gamma_{b-ns}^{ex} = -\partial E_{b-ns}^{ex}/\partial \varepsilon_{ext}$ and $\gamma_b^{tr} = -\partial E_b^{tr}/\partial \varepsilon_{ext}$ describe the changes in the exciton and trion

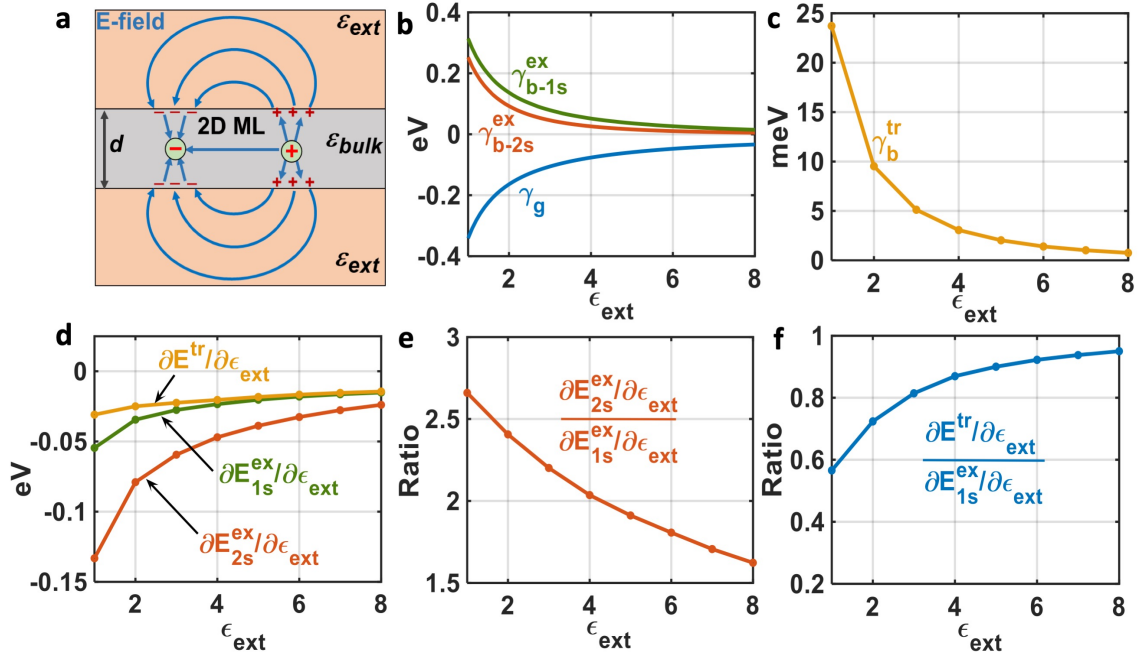

**Fig. 3. Theoretical model. a**, Dielectric polarization charge, which renormalizes the bandgap of a TMD monolayer, is depicted for the case $\varepsilon_{bulk} > \varepsilon_{ext}$. **b-c**, The coefficients $\gamma_g = \partial E_g / \partial \varepsilon_{ext}$, $\gamma_{b-1s}^{ex} = -\partial E_{b-1s}^{ex}/\partial \varepsilon_{ext}$, and $\gamma_{b-2s}^{ex} = -\partial E_{b-2s}^{ex}/\partial \varepsilon_{ext}$ (**b**) and $\gamma_b^{tr} = -\partial E_b^{ex}/\partial \varepsilon_{ext}$ (**c**) that describe bandgap renormalization, and the sensitivities of the binding energies of the *2s* and the *1s* exciton levels and the trion level, respectively, are plotted as functions of $\varepsilon_{ext}$. **d**, The sensitivities of the energies of the *2s* and the *1s* exciton levels and the trion level, including bandgap renormalization, are plotted as a function of $\varepsilon_{ext}$. **e-f**, The relative sensitivities of the energies of the *2s* and *1s* exciton levels (**e**), and of the trion and *1s* exciton levels (**f**), with respect to changes in $\varepsilon_{ext}$.

*binding* energies, respectively, due to dielectric disorder, and $\Delta\varepsilon_{ext}(\vec{R})$ represents the variation in the (relative) dielectric constant of the media surrounding the monolayer [10]. Note that $\varepsilon_{ext}$ is the average of the dielectric constants of the media on the top and bottom sides of the monolayer. The value of $\gamma_g$ can be obtained as the change in the energy of a hole due to dielectric polarization charges (Fig. 3a) in a thin film of thickness $d$, of bulk dielectric constant $\varepsilon_{bulk}$, and surrounded by medium of dielectric constant $\varepsilon_{ext}$, when $\varepsilon_{ext}$ changes by a small amount [20],

$$\gamma_g = \frac{\partial}{\partial \varepsilon_{ext}} \int \frac{d^2\vec{q}}{(2\pi)^2} \frac{e}{2\varepsilon_o q} \left[ \frac{1}{\varepsilon_{2D}(q)} - \frac{1}{\varepsilon_{bulk}} \right] \quad (3)$$

where the dielectric constant $\varepsilon_{2D}(q)$ is given by [20],

$$\varepsilon(q) = \varepsilon_{bulk} \frac{(1+\varepsilon_{bulk}/\varepsilon_{ext})+(1-\varepsilon_{bulk}/\varepsilon_{ext})e^{-qd}}{(1+\varepsilon_{bulk}/\varepsilon_{ext})-(1-\varepsilon_{bulk}/\varepsilon_{ext})e^{-qd}} \tag{4}$$

The values of $\gamma_{b-ns}^{ex}$ and $\gamma_b^{tr}$ can be computed using methods discussed previously [20]. The results are shown in Fig. 3b-c. Consistent with a previous report [10], our calculations show that effects due to bandgap renormalization and exciton binding energy shift almost cancel each other for the *1s* exciton state such that $\left|\partial E_{2s}^{ex}/\partial\varepsilon_{ext}\right|$ for the *2s* exciton state is almost exactly a factor of two larger than $\left|\partial E_{1s}^{ex}/\partial\varepsilon_{ext}\right|$ when $\varepsilon_{ext} \sim 4$ (the dielectric constant of h-BN) (Fig. 3d and 3e). We also find that the trion energy closely tracks the *1s* exciton energy such that $\left|\partial E^{tr}/\partial\varepsilon_{ext}\right|$ is ~0.87 $\left|\partial E_{1s}^{ex}/\partial\varepsilon_{ext}\right|$ when $\varepsilon_{ext} \sim 4$ (Fig. 3d and 3f). Assuming that $\Delta E_g^{S,T}(\vec{R})$ and $\Delta\varepsilon_{ext}(\vec{R})$ are statistically independent, the following quantities can be obtained directly from the covariance matrices of our data given in Equation (1),

$$\frac{\partial E_{2s}^{ex}/\partial\varepsilon_{ext}}{\partial E_{1s}^{ex}/\partial\varepsilon_{ext}} = \frac{K_{12}^a(E_{1s}^{ex},E_{2s}^{ex})-K_{22}^a(E_{1s}^{ex},E_{2s}^{ex})}{K_{11}^a(E_{1s}^{ex},E_{2s}^{ex})-K_{12}^a(E_{1s}^{ex},E_{2s}^{ex})} = 1.91 \tag{5},$$

$$\frac{\partial E^{tr}/\partial\varepsilon_{ext}}{\partial E_{1s}^{ex}/\partial\varepsilon_{ext}} = \frac{K_{12}^b(E_{1s}^{ex},E^{tr})-K_{22}^b(E_{1s}^{ex},E^{tr})}{K_{11}^b(E_{1s}^{ex},E^{tr})-K_{12}^b(E_{1s}^{ex},E^{tr})} = 0.89 \tag{6}.$$

The experimentally determined values of 1.91 and 0.89 for the ratios above are in remarkably agreement with the respective theoretical values of 2.03 and 0.87 (for $\varepsilon_{ext} \sim 4$). This agreement shows that the model given in Equation (2) captures the essential physics. In the Supplementary Material, we show that $\partial E_{2s}^{ex}/\partial E_{1s}^{ex}$ calculated in the case of potential disorder is given by the ratio of the polarizabilities of the *2s* and *1s* exciton states and equals ~102, which is ~53 times larger than the measured value of 1.91. We therefore conclude that potential disorder is not the main contributor to the variations in exciton and trion energies in our samples.

Based on equation (2), one can also use the covariance matrices to obtain root mean square values $\sqrt{\left\langle\left(\Delta E_g^{S,T}\right)^2\right\rangle}$ and $\sqrt{\left\langle\left(\Delta\varepsilon_{ext}\right)^2\right\rangle}$ from the following relations,

$$\sqrt{\left\langle \left(\Delta \varepsilon_{ext}\right)^2 \right\rangle} = \frac{1}{\left|\partial E_{1s}^{ex}/\partial \varepsilon_{ext}\right|} \sqrt{\frac{\left(K_{11}^{a/b} - K_{12}^{a/b}\right)^2}{K_{11}^{a/b} + K_{22}^{a/b} - 2K_{12}^{a/b}}}$$

$$\sqrt{\left\langle \left(\Delta E_g^{S,T}\right)^2 \right\rangle} = \sqrt{\frac{K_{11}^{a/b} K_{22}^{a/b} - \left(K_{12}^{a/b}\right)^2}{K_{11}^{a/b} + K_{22}^{a/b} - 2K_{12}^{a/b}}}$$
(7)

Using the theoretical value ~23.5 meV of $\left|\partial E_{1s}^{ex}/\partial \varepsilon_{ext}\right|$ (Fig. 3d), we find that $\sqrt{\left\langle \left(\Delta \varepsilon_{ext}\right)^2 \right\rangle}$ equals 0.0813 if we use the covariance matrix $K^a\left(E_{1s}^{ex}, E_{2s}^{ex}\right)$ and 0.0810 if we use the covariance matrix $K^b\left(E_{1s}^{ex}, E^{tr}\right)$. This remarkable agreement between the values of $\sqrt{\left\langle \left(\Delta \varepsilon_{ext}\right)^2 \right\rangle}$ obtained using two different experimental techniques (PL and reflection spectroscopies) that looked at two different energy level differences (between *2s* and *1s* exciton levels in the case of reflection spectroscopy and between *1s* exciton and trion levels in the case of PL) further supports the validity of our theoretical model. The values of $\sqrt{\left\langle \left(\Delta E_g^{S,T}\right)^2 \right\rangle}$ come out to be 2.4 meV and 2.9 meV if we use the covariance matrices $K^a\left(E_{1s}^{ex}, E_{2s}^{ex}\right)$ and $K^b\left(E_{1s}^{ex}, E^{tr}\right)$, respectively. These $\sqrt{\left\langle \left(\Delta \varepsilon_{ext}\right)^2 \right\rangle}$ values are in satisfactory agreement given the very different experimental measurements.

In conclusion, we have compared the behaviors of excitons and trions in the presence of disorder in the state-of-the-art sample of monolayer MoSe$_2$ encapsulated in h-BN. Hyperspectral imaging revealed that the *2s-1s* exciton energy splitting varies by $\sigma_\Delta^{ex}$=1.7±0.1 meV due to disorder. In contrast, the trion binding energy is robust with spatial variation of only $\sigma_b^{tr}$= 0.186±0.001 meV, which is one order of magnitude lower than $\sigma_\Delta^{ex}$. Theoretical analysis based on the many-body exciton-trion quantum superposition model [20] provides quantitative explanation of the experimental results.

**ACKNOWLEDGEMENT**. The experimental work was supported by the Materials Science and Engineering Research Center (MRSEC) through NSF grant DMR-2011738. Sample preparation was supported by the Vannevar Bush Faculty Fellowship program through Office of Naval

Research Grant # N00014-18-1-2080. We thank Kenji Watanabe and Takashi Taniguchi for providing h-BN crystals and Wenjing Wu, Lin Zhou, and Song Liu for help with sample fabrication. The theoretical work was supported by CCMR under NSF-NRSEC grant number DMR-1719875, and from NSF EFRI-NewLaw under grant number 1741694, and from AFOSR under Grant FA9550-19-1-0074.

**Supplementary Materials**

Methods, Additional Data and Analysis. Figures S1 to S2.

# Supplementary Material

**Disorder of Excitons and Trions in Monolayer MoSe$_2$**

Jue Wang[1], Christina Manolatou[2], Yusong Bai[1], James Hone[3], Farhan Rana[2,*], and X.-Y. Zhu[1,*]

[1] Department of Chemistry, Columbia University, New York, NY 10027, USA

[2] School of Electrical and Computer Engineering, Cornell University, Ithaca, NY 14853, USA

[3] Department of Mechanical Engineering, Columbia University, New York, NY 10027, USA

## Methods

<u>Sample preparation.</u> MoSe$_2$ monolayers was mechanically exfoliated from bulk crystals grown by the self-flux method with low defect densities (< $10^{11}$ cm$^{-2}$). h-BN flakes of thicknesses 10 to 50 nm were exfoliated and used to encapsulate the MoSe$_2$ monolayer by the polymer-free van der Waals assembly technique. A transparent polydimethylsiloxane stamp coated by a thin layer of polypropylene carbonate (PPC) was used to pick up an h-BN. This h-BN was used to pick up the MoSe$_2$ monolayer and then stamped onto a second h-BN and detached from the PPC at elevated temperatures (90° to 120°C). The residual PPC was washed away by acetone to give a clean h-BN/MoSe$_2$/h-BN heterostructure on the Si/SiO$_2$ substrate.

<u>Optical setup.</u> All spectroscopy measurements were performed on a home-built scanning confocal microscope system based on a liquid-helium cryostation (Montana Instruments Fusion/X-Plane) with a 100×, NA 0.75 objective (Zeiss LD EC Epiplan-Neofluar 100×/0.75 HD DIC M27). In mapping experiments, the laser excitation and PL collection are scanned together across the sample; in evenly illuminated PL and PL diffusion experiments, the laser excitation was fixed while the collection was scanned. In steady-state PL measurements, a continuous wave laser (532 nm) was focused by the objective to a diffraction-limited spot on the sample. The PL was detected by an InGaAs photodiode array (PyLoN-IR, Princeton Instruments).

<u>Photoluminescence (PL) spectroscopy</u>**.** For PL spectroscopy, a continuous wave laser at 532 nm (2.33 eV) is used to excite the sample in a diffraction-limited FWHM focus spot of ~0.43 μm (using a 0.75 NA objective) with a steady-state light intensity of 4.66 μW/μm$^2$.

Reflectance spectroscopy. To obtain the spatial variations of the energy splitting between the 2s and 1s exciton levels and make a comparison to the trion binding energy, reflectance mapping was conducted on the BN encapsulated monolayer MoSe$_2$ sample. After flat-field correlation, reflectance contrast spectra is calculated as $R_C = \frac{R_{sample}}{R_{substrate}} - 1$. The energy of A 1s and A 2s excitons are then extracted empirically as follows. For A 1s, the absolute value of derivative of $R_C$ was baseline corrected and used as weight in averaging the energy around the resonance peak. For A 2s, the spectrum around resonance was fit to the sum of a quadratic background and a derivative of Gaussian function, the zero-crossing of which is assigned to A 2s energy. Such procedure is conducted for each spot in reflectance mapping to reconstruct the image and histogram of A 1s, A 2s energy and their difference.

Computational details. In computations, we used the following expression for the (relative) dielectric constant of an encapsulated 2D monolayer [1],

$$\varepsilon(q) = \varepsilon_{bulk} \frac{1 - \frac{(1-\varepsilon_{bulk}/\varepsilon_1)(1-\varepsilon_{bulk}/\varepsilon_2)}{(1+\varepsilon_{bulk}/\varepsilon_1)(1+\varepsilon_{bulk}/\varepsilon_2)} e^{-2qd}}{\left[1 - \frac{(1-\varepsilon_{bulk}/\varepsilon_1)}{(1+\varepsilon_{bulk}/\varepsilon_1)} e^{-qd}\right]\left[1 - \frac{(1-\varepsilon_{bulk}/\varepsilon_2)}{(1+\varepsilon_{bulk}/\varepsilon_2)} e^{-qd}\right]} \tag{s1}$$

Here, $\varepsilon_1$ and $\varepsilon_2$ are the dielectric constants of the media above and below the monolayer and $\varepsilon_{bulk}$ is the bulk dielectric constant of the material making up the monolayer. In the limit $qd \to 0$, the above expression reduces to the standard Keldysh form [2],

$$\varepsilon(q) = \left(\frac{\varepsilon_1 + \varepsilon_2}{2}\right)(1 + r_o q) \tag{s2}$$

In the case $\varepsilon_1 = \varepsilon_2 = \varepsilon_{ext}$, the above expression simplifies to,

$$\varepsilon(q) = \varepsilon_{bulk} \frac{(1+\varepsilon_{bulk}/\varepsilon_{ext}) + (1-\varepsilon_{bulk}/\varepsilon_{ext})e^{-qd}}{(1+\varepsilon_{bulk}/\varepsilon_{ext}) - (1-\varepsilon_{bulk}/\varepsilon_{ext})e^{-qd}} \tag{s3}$$

For MoSe$_2$, we have used a value of 12 for $\varepsilon_{bulk}$ [3]. We have assumed a values of $0.7 m_o$ for the effective masses of both electrons and holes in MoSe$_2$ [3]. We assume variational expressions for the exciton 1s, 2s, and 2p wavefunctions with forms similar to the standard 2D exciton wavefunctions and use the radii as the variational parameters [1–3]. The computational details are given in a previous work [3]. We also used a variational expression for the lowest trion state and used the two trion radii as variational parameters [3].

**Additional data and analysis**

<u>Temperature dependent PL spectra.</u>  The insensitivity of the trion binding energy is seen in most of the temperature range where the trion is observed (Fig. S1). Figs. S2a-c show the histograms of $E_{1s}^{ex}$, $E^{tr}$, and $E_b^{tr}$ at different temperatures (T). Both $E_{1s}^{ex}$ and $E^{tr}$ are broadly distributed due to disorder at all temperatures, while the distribution of $E_b^{tr}$ remains very narrow until the temperature exceeds ~60K when the trion PL intensity becomes negligible (Fig. S1). The observed temperature-dependent redshifts in $E_{1s}^{ex}$ and $E^{tr}$ (Fig. S2d) are in agreement with previous report by Ross et al. [4]. The red shift in the energies of the exciton and trion lines with temperature is attributed to the temperature-dependent reduction in the material bandgap [5]. Since the trion binding energy is dependent on the sample electron density [3], a slight increase in the sample electron density with the temperature can explain the increase in the trion binding energy with temperature (Fig. S2e). While the trion PL intensity ($I^{tr}$) is higher than that of the exciton ($I_{1s}^{ex}$) at low temperatures, $I^{tr}$ falls below $I_{1s}^{ex}$ at T ~60K and disappears above this temperature (Fig. S1), an observation previously attributed to the thermal dissociation of the trion [4]. At T ~ 60 K, the trion binding energy matches the energy of the $A_1'$ out-of-plane transverse optical phonon mode (Fig. S1), suggesting the involvement of strong phonon scattering in the dissociation of the trion. The width $\sigma_b^{tr}$ of the distribution of the trion binding energies, Fig. S2f, remains small and nearly constant until ~60 K when the resonant condition is reached and phonon scattering becomes important.

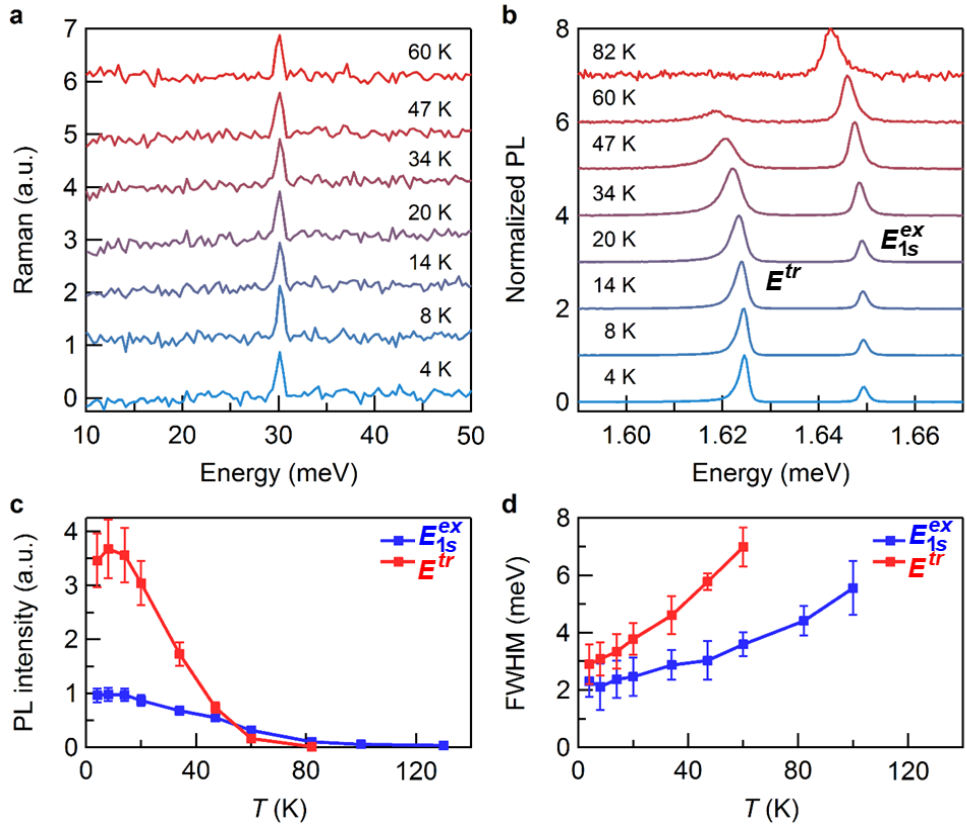

**Fig. S1. Temperature dependence of PL and Raman spectra. a**, Temperature dependence of Raman spectrum of monolayer MoSe$_2$ showing a nearly constant A$_1$' peak from 4 K to 60 K. The background Raman scattering of BN on SiO$_2$/Si substrate is subtracted. **b**, Temperature dependent PL spectrum of a representative spot of the h-BN encapsulated monolayer MoSe$_2$. **c** and **d**, PL intensity (**c**) and FWHM linewidth (**d**) of exciton and trion peaks as functions of temperature. Error bars represent widths of intensity and FWHM linewidth distributions across the sample.

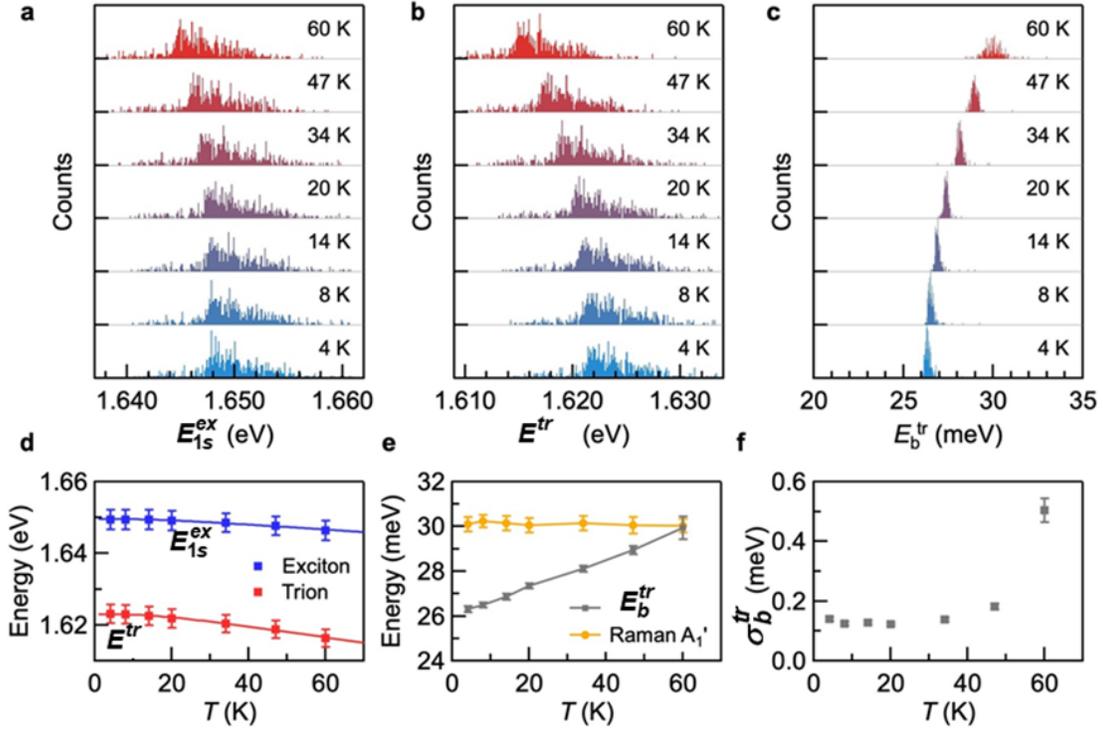

**Fig. S2. Temperature dependence of trion binding in monolayer MoSe$_2$. a-c**, Temperature dependent histograms of exciton peak energy $E_{1s}^{ex}$ (**a**), trion peak energy $E^{tr}$ (**b**) and trion binding energy $E_b^{tr}$ (**c**) with a uniform energy span of 25 meV. (**d**) Peak exciton and trion energies as a function of temperature. (**e**) A comparison between the temperature dependent trion binding energy $E_b^{tr}$ and the A1' phonon energy; note the resonance condition at ~ 60K. (**f**) Temperature dependence of the distribution of the trion binding energies. Note the abrupt increase in the width of the distribution around ~60K.

The case against potential disorder. In this Section we show that potential disorder is inconsistent with our data. We assume that the potential disorder is described by a slowly varying potential energy function $V$. We calculate the shift in the exciton and trion energies cause by disorder. Suppose the wavefunction of an exciton state in the relative coordinate $\vec{r} = \vec{r}_e - \vec{r}_h$ is $\psi_\alpha^{ex}(\vec{r})$. The subscript $\alpha$ refers to *1s*, *2s*, *2p*, etc., the disorder potential experienced by the exciton can be written to first order as,

$$V(\vec{R} - \lambda_h \vec{r}) - V(\vec{R} - \lambda_e \vec{r}) \approx \vec{r} \cdot \nabla_{\vec{R}} V(\vec{R}) + \ldots \ldots \quad (s4)$$

where, $\vec{R}$ is the center of mass coordinate of the exciton, and $\lambda_e = 1 - \lambda_h = m_e/m_{ex} \approx 1/2$. $m_e$ ($m_h$) is the electron (hole) effective mass, and $m_{ex} = m_e + m_h$. Since the exciton is charge neutral, only the gradient of the disorder potential $V(\vec{R})$, and not $V(\vec{R})$ itself, appears to first order in the expressions above. The local variations in the exciton energies are found using perturbation theory,

$$\Delta E_{1s}^{ex}(\vec{R}) \approx -\left(\frac{\sqrt{3}}{2}\right)^2 \left(\frac{16\left(a_{1s}^{ex} a_{2p}^{ex}\right)^2}{\left(a_{1s}^{ex} + a_{2p}^{ex}\right)^4}\right)^2 \frac{\left|\nabla_{\vec{R}} V(\vec{R}) a_{1s}^{ex}\right|^2}{\left(E_{2p}^{ex} - E_{1s}^{ex}\right)} \quad (s5)$$

$$\Delta E_{2s}^{ex}(\vec{R}) \approx -\left(\frac{8\left(a_{2s}^{ex} a_{2p}^{ex}\right)^2}{\left(a_{2s}^{ex} + a_{2p}^{ex}\right)^5}\left(7a_{2p}^{ex} - a_{2s}^{ex}\right)\right)^2 \frac{\left|\nabla_{\vec{R}} V(\vec{R}) a_{2s}^{ex}\right|^2}{\left(E_{2p}^{ex} - E_{2s}^{ex}\right)} \quad (s6)$$

Here, $a_{1s}^{ex}$, $a_{2s}^{ex}$, and $a_{2p}^{ex}$ are the radii of the *1s*, *2s*, and *2p* exciton states. Note that the energies $E_{2p}^{ex}$ and $E_{2s}^{ex}$ of the *2p* and *2s* exciton states are not degenerate in 2D TMD materials[25]. The terms on the left-hand side represent the perturbative changes in the exciton energies as a result of Stark shifts caused by the local electric field which is proportional to the gradient of the disorder potential. The electric field polarizes the lowest energy *1s* and *2s* exciton states by mixing exciton states with *s* and *p* symmetries. The electric field polarizability $\chi_\alpha$ of an exciton state is defined by the relation [6], $\Delta E_\alpha^{ex} \propto 0.5 \chi_\alpha |F|^2$, where $F$ is the electric field strength. Using the expression given in Equation (s5), the polarizability of the *1s* exciton state in h-BN encapsulated MoSe$_2$ is found to be ~$8.5 \times 10^{-37}$ J-m$^2$/V$^2$ in very good agreement with the value of ~$10.1 \times 10^{-37}$ J-m$^2$/V$^2$ reported previously [6]. In comparison, the polarizability of the *2s* exciton state, found using Equation (s6), is ~$8.67 \times 10^{-35}$ J-m$^2$/V$^2$, which is almost two orders of magnitude larger. The value of the ratio $\left(\partial E_{2s}^{ex}/\partial F^2\right)/\left(\partial E_{1s}^{ex}/\partial F^2\right)$, which is equal to the ratio of the polarizabilities $\chi_{2s}/\chi_{1s}$, is ~102 which is much larger than then experimentally found value of ~1.91. We therefore conclude that potential disorder is not the main contributor to the disorder we observe in our samples.

**References for supplementary information**